\documentclass[prd,aps,floatfix,preprintnumbers,groupedaddress,nofootinbib,reprint]{revtex4-1}
\pdfoutput=1

\usepackage{graphicx, url}
\usepackage{color}
\usepackage{amsmath, amssymb,bbm}

\begin{document}   
\title{Probing Left-Right Seesaw using Beam Polarization at an $e^+e^-$ Collider} 
\author{Sudhansu S. Biswal$^1$ and P. S. Bhupal Dev$^2$}
\affiliation{$^1$ Department of Physics, Ravenshaw University, Cuttack 753003, India\\
$^2$ Department of Physics and McDonnell Center for the Space Sciences, Washington University, St.~Louis, MO 63130, USA}
\begin{abstract}
We show that the longitudinal beam polarization option at a future electron-positron collider provides an unambiguous distinction between low-scale seesaw models of neutrino mass. This is possible due to the fact that the pair production cross section of the heavy neutrinos in seesaw models is sensitive to the polarization of the initial lepton beams, and for a suitable choice of the polarization, shows a clear enhancement over the unpolarized cross section. More interestingly, the choice of the beam polarization for which the enhancement is maximum is governed by the size of the light-heavy neutrino mixing parameter. We also find that using this effect, one can probe a previously uncharted parameter space of the left-right seesaw model, which is complementary to the existing searches at both energy and intensity frontiers. 
\end{abstract}
\maketitle
\section{Introduction}
The seesaw mechanism provides an elegant and natural way to understand the observed smallness of 
neutrino masses. The simplest version, known as the type-I seesaw~\cite{type1a,type1b,type1c,type1d,type1e}, requires the existence of heavy right-handed (RH) neutrinos with Majorana masses. Although the mass scale of these RH neutrinos is {\it a priori} unknown, the possibility of having them close to the TeV scale has received much attention in recent years, as this scenario could be directly tested in the ongoing Large Hadron Collider (LHC) experiments~\cite{Keung:1983uu, Deppisch:2015qwa}, as well as indirectly probed in various low-energy searches for rare lepton number violation (LNV) and/or lepton flavor violation (LFV)~\cite{deGouvea:2013zba}. There are also theoretical arguments based on naturalness of the observed 125 GeV Higgs mass, which suggest the seesaw scale to be below $\sim 10^7$ GeV~\cite{Vissani:1997ys, Clarke:2015gwa, Bambhaniya:2016rbb}, thereby making the case of low-scale seesaw more compelling.

As far as the collider signals are concerned, in the minimal seesaw with just the Standard Model (SM) particles plus the RH neutrinos (denoted henceforth as the SM seesaw), the production of the RH neutrinos crucially depends on the size of the Dirac Yukawa coupling $Y_\nu$, defined by the Lagrangian 
\begin{align}
-{\cal L}_Y \ = \ Y_\nu \overline{N}\phi \psi_L+\frac{1}{2}\overline{N}M_N N^c+ {\rm H.c.} \, ,
\label{eq:lagy}
\end{align}
where $\phi$ is the SM Higgs doublet, $\psi_L=(\nu, \ell)_L$ (with $\ell=e,\mu,\tau$) is the $SU(2)_L$ lepton doublet, $M_N$ generically denotes the mass of the SM-singlet RH neutrinos $N$, and the superscript $c$ stands for charge conjugation. Here, the key parameter is the light-heavy neutrino mixing $V_{\ell N}\simeq M_DM_N^{-1}$, where $M_D=vY_\nu$ is the Dirac neutrino mass matrix and $v$ is the electroweak vacuum expectation value, since the production cross section of the RH neutrinos at colliders is proportional to powers of $|V_{\ell N}|^2$, depending on the production channel~\cite{Datta:1993nm, Panella:2001wq, Han:2006ip, Bray:2007ru, delAguila:2007qnc, Atre:2009rg, 
Dev:2013wba, Alva:2014gxa, Das:2015toa}. The $\sqrt s=8$ TeV LHC searches have put upper limits on $|V_{\ell N}|^2\lesssim 10^{-2}-1$ (with $\ell=e,\mu$) for heavy neutrino masses in the 100-500 GeV range~\cite{Khachatryan:2015gha, Aad:2015xaa, Khachatryan:2016olu}. Note that the smallness of the left-handed (LH) neutrino masses in the type-I seesaw formula $M_\nu\simeq M_DM_N^{-1}M_D^{\sf T}\lesssim 0.1~{\rm eV}$ usually requires $|V_{\ell N}|\lesssim 10^{-6}\sqrt{{\rm 100~GeV}/M_N}$, thus suppressing the prompt collider signals; however, there exist models in which cancellations in the seesaw matrix  can allow for large mixing values within reach of current and future collider searches~\cite{Pilaftsis:1991ug, Gluza:2002vs, Kersten:2007vk, Xing:2009in, He:2009ua, Adhikari:2010yt, Ibarra:2010xw, Mitra:2011qr, Dev:2013oxa}. 

A natural theoretical framework which provides a TeV-scale renormalizable theory of the seesaw mechanism is the Left-Right (L-R) symmetric extension of the SM~\cite{LR1,LR2,LR3}. The two essential ingredients of seesaw, namely, the existence of RH neutrinos and the seesaw scale, emerge naturally in this scenario -- the former as the parity gauge partners of the LH neutrinos, and the latter as the scale of parity restoration. This model predicts new gauge interactions for the RH neutrinos with the RH gauge bosons $W_R,Z'$, and therefore, new contributions to the collider signal of RH neutrinos~\cite{Keung:1983uu, Ferrari:2000sp, Nemevsek:2011hz, Das:2012ii, Chen:2013fna, Patra:2015bga, Mitra:2016kov}, which could be sizable for TeV-scale seesaw, independent of the size of the LH-RH neutrino mixing parameter $V_{\ell N}$. Using the smoking gun same-sign dilepton plus dijet ($\ell^\pm\ell^\pm jj$) channel, $W_R$ masses up to 3 TeV and $M_N$ masses up to 2 TeV have been probed at the $\sqrt s=8$ TeV LHC~\cite{Aad:2015xaa, Khachatryan:2014dka}. 

\begin{figure}[t!]
\includegraphics[width=4cm]{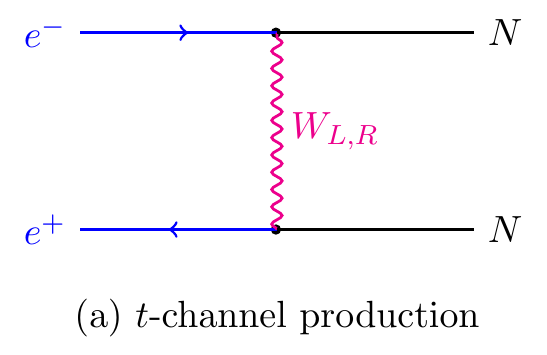} 
\includegraphics[width=4.5cm]{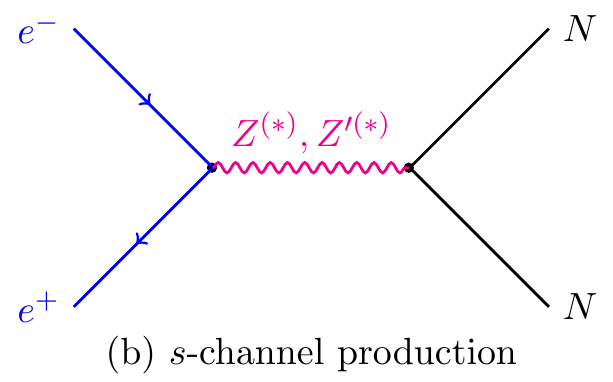} 
\includegraphics[width=3.9cm]{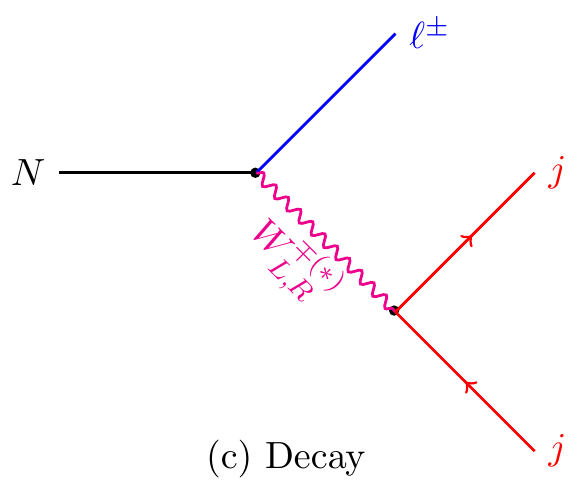}  
\caption{Feynman diagrams for the dominant pair-production and decay modes of the heavy neutrino at an $e^+e^-$ collider.}
\label{fig:1}
\end{figure}

In order to be able to directly probe the nature of the seesaw at colliders (i.e. whether it is SM seesaw or L-R seesaw, large mixing or small mixing, etc), it is important to distinguish between various contributions to a given RH neutrino signal. At the LHC, this is possible by studying various kinematics distributions~\cite{Chen:2013fna, Han:2012vk, Dev:2015kca}. In this paper, we present a new, simple way using the longitudinal beam polarization option at a future lepton collider, which generically applies to all existing proposals for an $e^+e^-$ machine, including ILC~\cite{Behnke:2013xla}, FCC-ee (formerly TLEP)~\cite{Gomez-Ceballos:2013zzn}, CLIC~\cite{Aicheler:2012bya}, CEPC~\cite{CEPC:2015csa}, and even to muon colliders~\cite{Kaplan:2014xda} (by just replacing $e \leftrightarrow \mu$ in our subsequent discussion). In particular, we show that with a suitable choice of the beam polarization, we can distinguish between LH and RH-current interactions by just measuring the total cross section of the pair-production process $e^+e^-\to NN$, followed the RH neutrino decay $N\to \ell^\pm jj$ (see Figure~\ref{fig:1}). What is more remarkable is the fact that using this process, we can probe $W_R$ masses up to 6.6 TeV  with a modest $500~{\rm fb}^{-1}$ of integrated luminosity at $\sqrt s=500$ GeV, and up to 9.4 TeV with $\sqrt s=1$ TeV, which extend the testability of L-R seesaw to an unprecedented level well beyond the LHC reach, even with its most optimistic high-luminosity (HL) upgrade.     

The rest of the paper is organized as follows: in Section~\ref{sec:2}, we make a comparative study of the unpolarized and polarized cross sections of our proposed signal at a generic $e^+e^-$ collider. In Section~\ref{sec:3}, we show the enhancement in the sensitivity of RH gauge boson searches by using the polarization effect. Our conclusions are given in Section~\ref{sec:4}. 
 
\section{Comparison between Unpolarized and Polarized Cross Sections}\label{sec:2}
We consider the heavy neutrino pair-production process $e^+e^-\to NN$ in the context of L-R seesaw (see Fig.~\ref{fig:1}). For earlier studies of various heavy neutrino signals at $e^+e^-$ colliders (including the $e^-e^-$ option), see e.g.~Refs.~\cite{Buchmuller:1991tu, Djouadi:1993pe, Azuelos:1993qu, Gluza:1995ix, Belanger:1995nh, Gluza:1995js, Ananthanarayan:1995cn, Gluza:1997ts, Achard:2001qv, Rodejohann:2010jh, Antusch:2015mia, Banerjee:2015gca, Asaka:2015oia, Wang:2016eln, Antusch:2016ejd} in the context of SM seesaw and Refs.~\cite{Huitu:1997vh, Raidal:2000ru, Barry:2012ga} in the context of L-R seesaw. However, the distinct advantage of the process considered here under beam polarization effects (as will become apparent below) has not been discussed before to the best of our knowledge.  For relatively large light-heavy neutrino mixing of the electron flavor $V_{eN}$, the dominant contribution to the production cross section of the $N$-pair comes from the $t$-channel $W_L$ exchange [cf.~Fig.~\ref{fig:1}(a)], with a sub-dominant contribution from the $s$-channel (off-shell) $Z$ exchange (for $M_N>M_Z/2$) [cf.~Fig.~\ref{fig:1}(b)]. On the other hand, for small $V_{eN}$, the $W,Z$-mediated processes are suppressed by $|V_{eN}|^4$. In this case, for TeV-scale RH gauge boson masses, the $W_R$-mediated $t$-channel process in Fig.~\ref{fig:1}(a) becomes dominant, its cross section being independent of $V_{eN}$ and only suppressed by $(m_{W}/M_{W_R})^4$. There is also a sub-leading contribution from the $s$-channel exchange of off-shell $Z'$, as shown in Fig.~\ref{fig:1}(b). Both $s$ and $t$-channel contributions are included in our numerical calculation. There are additional contributions from the $s$-channel exchange of the SM Higgs boson, as well as other heavy neutral scalars present in L-R seesaw; however, these contributions are negligible mainly due to their tiny Yukawa coupling to electrons. 

Similarly, for the decay of the heavy neutrino, depending on the size of $V_{\ell N}$ and $M_{W_R}$, either the LH or RH-current processes could dominate. For concreteness, let us assume a single heavy neutrino mass eigenstate which is dominantly of electron flavor, so that the mixing parameters $V_{\mu N}$ and $V_{\tau N}$ can be safely ignored in the decay rate calculation. In this simplistic scenario, the LFV constraints from rare decays such as $\mu\to e\gamma$ are not applicable. For large $V_{eN}$, the two-body processes $N\to W_L^\mp e^\pm,~Z\nu(\bar{\nu}),~H\nu(\bar{\nu})$ (where $H$ is the SM Higgs boson), respectively mediated by the LH charged-current and neutral-current and Dirac Yukawa interactions, are dominant, provided they are kinematically allowed. The respective partial decay widths are given by 
\begin{align}
\Gamma(N\to We) & = \frac{g^2M_N^3}{64\pi m_W^2}|V_{eN}|^2\left(1-\frac{m^2_{W}}{M^2_N}\right)^2\left(1+\frac{2m_W^2}{M_N^2}\right),\nonumber \\
\Gamma(N\to Z\nu) & = \frac{g^2M_N^3}{128\pi m_W^2}|V_{eN}|^2\left(1-\frac{m^2_{Z}}{M^2_N}\right)^2\left(1+\frac{2m_Z^2}{M_N^2}\right),\nonumber \\
\Gamma(N\to H\nu) &=\frac{g^2M_N^3}{128\pi m_W^2}|V_{eN}|^2\left(1-\frac{m^2_{H}}{M^2_N}\right)^2,
\label{eq:2body}
\end{align}
where $g$ is the $SU(2)_L$ gauge coupling strength. For $M_N\gg m_W$, the ratio of the decay rates in Eqs.~\eqref{eq:2body} becomes $\Gamma(N\to We):\Gamma(N\to Z\nu):\Gamma(N\to H\nu)\simeq 2:1:1$. For $M_N<m_W$, none of the above 2-body decays is kinematically allowed, and one should instead consider the 3-body processes with off-shell gauge and Higgs boson exchange. For the analytic expressions of the relevant 3-body partial decay widths, see e.g.~Refs.~\cite{Atre:2009rg, Helo:2010cw}. 

\begin{figure*}[ht]
\includegraphics[width=8.5cm]{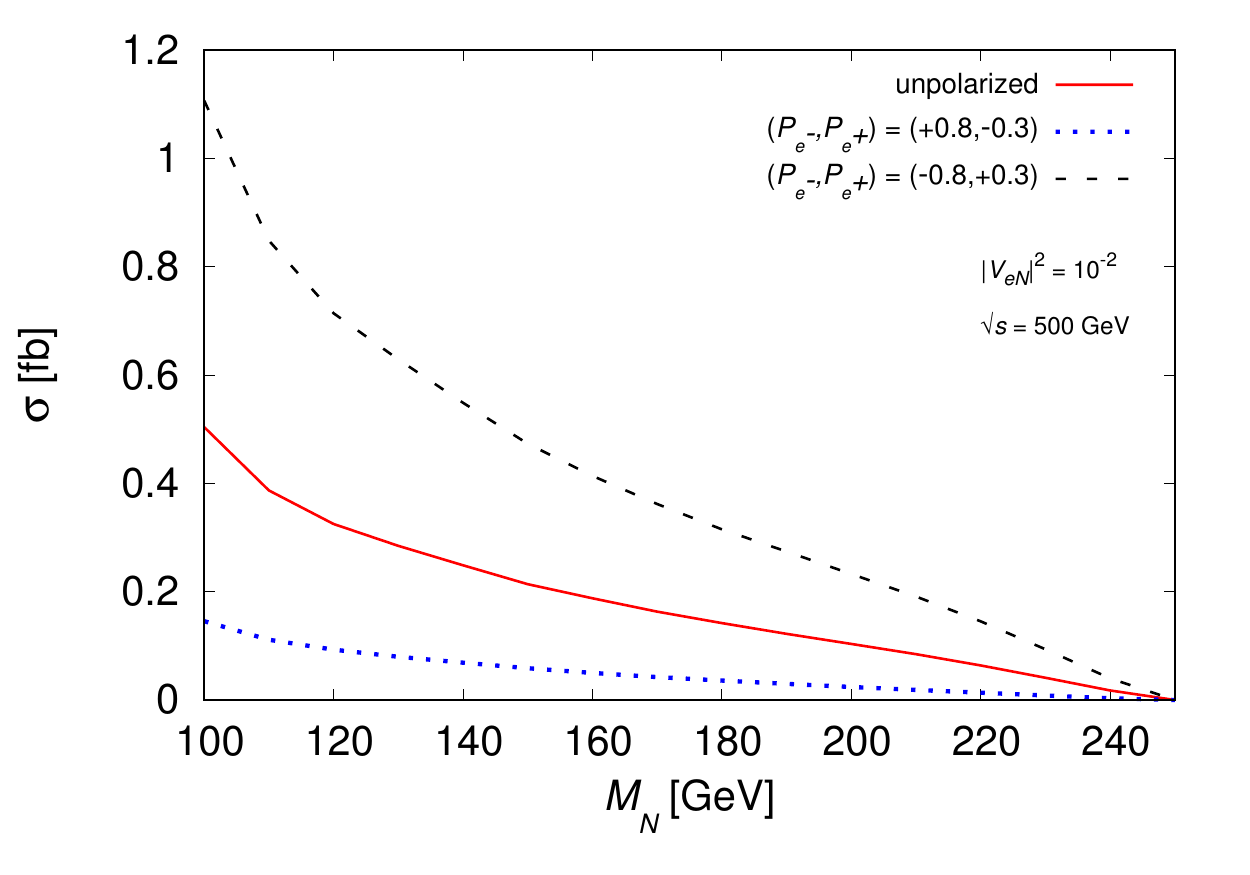}
\includegraphics[width=8.5cm]{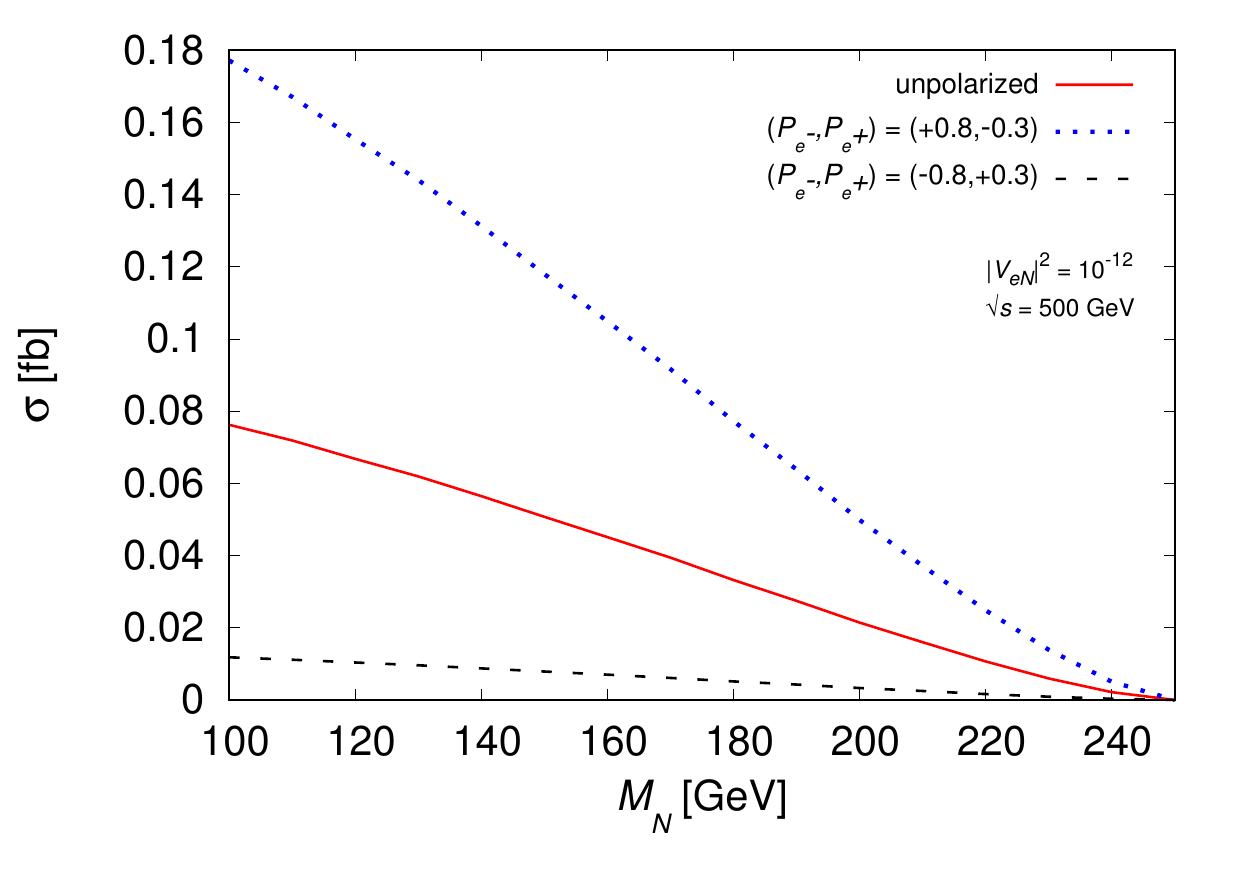}
\caption{Comparison between the unpolarized and polarized cross sections for the process $e^+e^-\to NN\to e^\pm e^\pm+4j$ at $\sqrt s=500$ GeV with two different benchmark values of $|V_{eN}|^2=10^{-2}$ ({\it left panel}) and $10^{-12}$ ({\it right panel}). Here we have taken $M_{W_R}=3$ TeV and $g_R=g_L$ for illustration. In each panel, we consider two realistic choices for the longitudinal beam polarization: $\pm 80\%$ for the electron and $\mp 30\%$ for the positron.}
\label{fig:2}
\end{figure*}

In L-R seesaw, one should also consider the 3-body decay $N\to W_R^{\pm*}e^\mp \to e^\mp jj$ mediated by purely RH charged-current interaction, with the partial width
\begin{align}
\Gamma(N\to W_R^*e \to ejj) \simeq \frac{3g_R^4}{2048\pi^3}\frac{M_N^5}{M_{W_R}^4} ,
\label{eq:3body}
\end{align}
where $g_R$ is the $SU(2)_R$ gauge coupling strength and we have assumed $M_N\ll M_{W_R}$, as favored by vacuum stability considerations~\cite{Mohapatra:1986pj, Maiezza:2016bzp}. Thus, for on-shell production of $N$ pairs, taking the narrow-width approximation, the total cross section of our proposed signal is given by 
\begin{align}
& \sigma(e^+e^-\to NN\to e^\pm e^\pm+4j) \nonumber \\
& \qquad = \sigma(e^+e^-\to NN)\times [{\rm BR}(N\to e^\pm jj)]^2 ,
\label{eq:signal}
\end{align}
where ${\rm BR}(N\to ejj)=\Gamma(N\to ejj)/\Gamma_N$ and $\Gamma_N$ is the total $N$ decay width. When the 2-body decays of $N$ are dominant,  $\Gamma(N\to ejj)=\Gamma(N\to We){\rm BR}(W\to jj)\simeq 0.676\: \Gamma(N\to We)$. 
For heavy Majorana neutrino decays, the signal will consist of 50\% same-sign and  50\% opposite-sign dilepton events, both with 4 additional jets. The same-sign dilepton events suffer from much less SM background (mainly due to charge misidentification), as compared to the opposite-sign dilepton events, and therefore, we will focus on the same-sign events. But it is worth noting that even if we consider only the opposite-sign dilepton events, which might indeed be the case when the same-sign dilepton signal is suppressed, as e.g. in low-scale SM seesaw models~\cite{Kersten:2007vk, Ibarra:2010xw} or in some variants of the type-I seesaw~\cite{Mohapatra:1986aw, Mohapatra:1986bd, Malinsky:2005bi, Dev:2012sg} with approximate lepton number conservation, the main new result of this section, namely, the effect of polarization on the total cross section remains unchanged. In this sense, the results of this section are generically applicable to a wider variety of seesaw models.

Our numerical results  for the signal cross section given by Eq.~\eqref{eq:signal}, calculated using a {\tt FeynRules}~\cite{Alloul:2013bka} implementation of the L-R seesaw model~\cite{Roitgrund:2014zka} in {\tt MadGraph5}~\cite{Alwall:2014hca}, are shown in Fig.~\ref{fig:2} for a modest value of center of mass energy $\sqrt s=500$ GeV. To see the effect of beam polarization on the total cross section, we have considered two benchmark cases: (i) large light-heavy neutrino mixing with $|V_{eN}|^2=10^{-2}$, which represents the SM seesaw case (with fine-tuning in the seesaw matrix), while satisfying the current experimental constraints from the LHC~\cite{Khachatryan:2016olu} and elsewhere~\cite{Deppisch:2015qwa}, and (ii) small light-heavy neutrino mixing with  $|V_{eN}|^2=10^{-12}$, which represents the minimal L-R seesaw case (without any fine-tuning in the seesaw matrix). For concreteness, we assume $g_R=g_L$ for the gauge couplings and choose a typical value of $M_{W_R}=3$ TeV which is consistent with current experimental bounds~\cite{Aad:2015xaa, Khachatryan:2014dka}. We have also imposed the basic trigger cuts $p_T^e>10$ GeV, $p_T^j>20$ GeV, $|\eta_e|<2.5$, $|\eta_j|<5$ and $\Delta R_{jj},\Delta R_{ej}>0.4$ on the final-state leptons and jets, following a generic ILC detector design~\cite{Behnke:2013lya}. We find that the cut efficiency for our signal is between 60-70\% for $M_N>m_W$ and drops rapidly for $M_N<m_W$. In any case, we obtain sizable cross sections for heavy neutrino pair-production, as long as it is kinematically allowed. 

Note that being a lepton collider, the $e^\pm e^\pm+4j$ signal enjoys a clean environment with virtually no SM background, except when the charge of one of the final state leptons in $e^+e^-+4j$ is misidentified. However, the charge misidentification rate in the ILC detector~\cite{Behnke:2013lya} is expected to be very small, and certainly below that of the LHC, which is at the level of $1\%$~\cite{Khachatryan:2015hwa}. Using {\tt MadGraph5}, we estimate the SM background for $e^+e^-+4j$ at leading order after applying the same cuts as above to be 0.14 fb, 0.10 fb and 0.22 fb for the unpolarized, $(+0.8,-0.3)$ and $(-0.8,+0.3)$ configurations, respectively. Taking a conservative value of 1\% for the charge misidentification rate at ILC, we infer that the SM background for our $e^\pm e^\pm+4j$ signal is negligible. Even in the worst case scenario when all four jets in the final state cannot be tagged, the total SM background for the corresponding inclusive process $e^\pm e^\pm+nj$ (with $n\geq 2$) is found to be of the same order as the signal cross section for the large mixing scenario.

For a given heavy neutrino mass $M_N$, the small mixing case is expected to have a smaller total cross section, simply because of the $(m_W/M_{W_R})^4\simeq 5\times 10^{-7}$ kinematic suppression, as compared to the large mixing case, which is only suppressed by a factor $|V_{eN}|^4=10^{-4}$ in our case, apart from the $\sim 50\%$ branching fraction of the $N\to e^\pm jj$ final state. But what is more interesting in Fig.~\ref{fig:2} is the {\it relative} change in the cross section when we switch on the beam polarization for the initial electron and positron beams, as compared to the corresponding unpolarized cross sections. In particular, for the large mixing case, since the heavy neutrino production is dominated by the $t$-channel $W_L$ exchange which is purely LH interaction,  we expect an enhancement (suppression) of the cross section by making the electrons left (right)-polarized and positrons right (left)-polarized. This effect is demonstrated in Fig.~\ref{fig:2} (left panel) by taking two realistic configurations with $\pm 80\%$ and $\pm 30\%$ polarization of the initial electron and positron beams, respectively. 

Now for the small mixing case, since the LH-current interactions are all suppressed by $|V_{eN}|^4$, the $t$-channel $W_R$ mediated process gives the dominant contribution to the signal. Here, we would expect just the opposite relative change in the polarized versus unpolarized cross sections, i.e.  an enhancement (suppression) of the cross section by making the electrons right (left)-polarized and positrons left (right)-polarized. This is exactly what happens in Fig.~\ref{fig:2} (right panel). Therefore, just measuring the total cross sections and comparing those with and without beam polarization at an $e^+e^-$ collider, one can effectively probe the heavy-light neutrino mixing parameter, which is an essential ingredient of the seesaw mechanism. In this way, one can unambiguously distinguish between different seesaw models. This is a generic result, applicable to all proposed designs for future $e^+e^-$ machines, irrespective of whether they are linear (e.g. ILC, CLIC) or circular (e.g., FCC-ee, CEPC). 

\section{Enhanced Sensitivity to $W_R$}\label{sec:3}
\begin{figure}[t!]
\includegraphics[width=8cm]{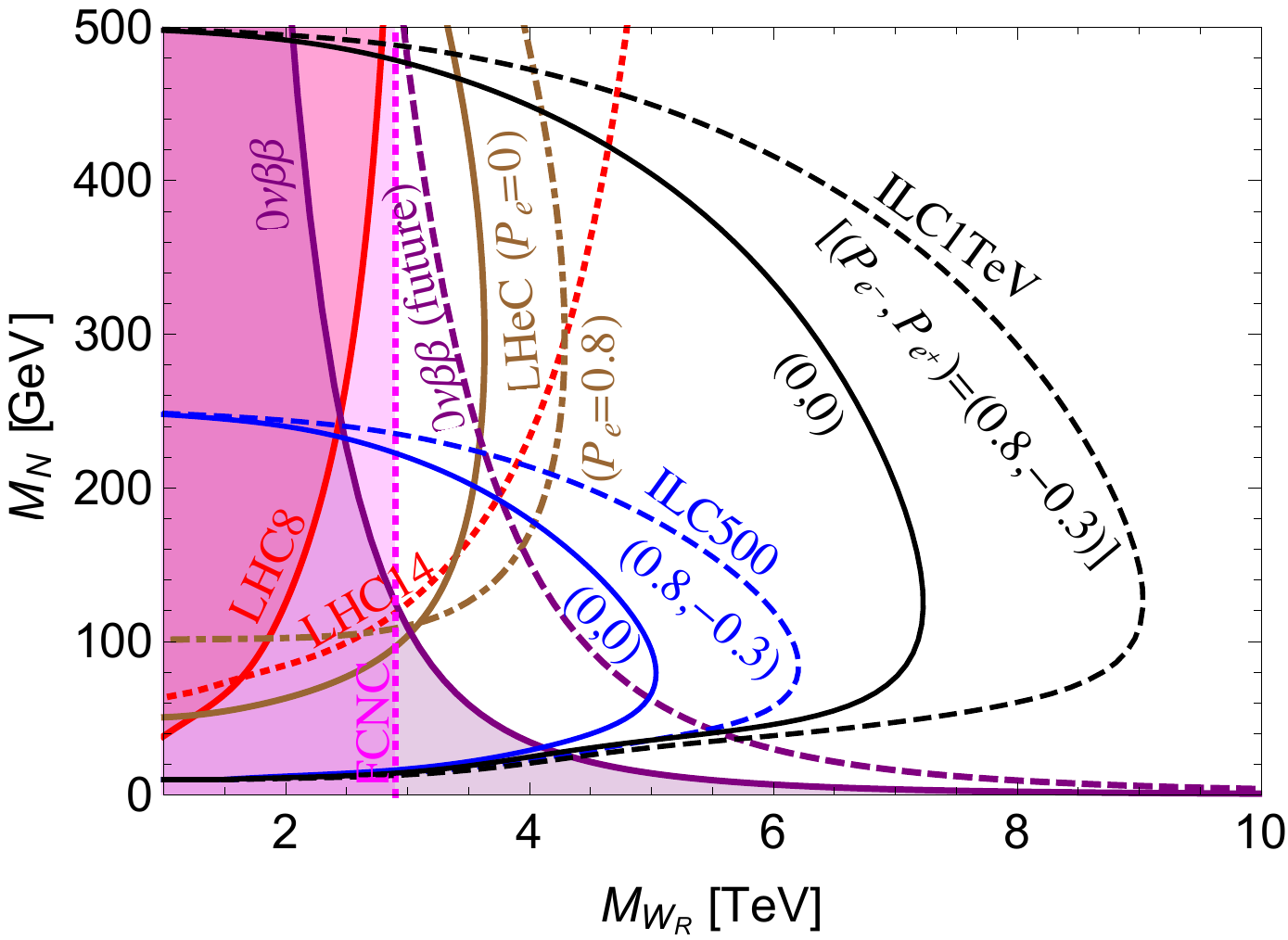}
\caption{95\% CL sensitivity contours in the $(M_{W_R},M_N)$ plane for our proposed signal at $\sqrt s=500$ GeV (blue) and 1 TeV (black) ILC, both without (solid) and including (dashed) polarization effects, as indicated by the values of $(P_{e^-},P_{e^+})$. For comparison, we also show the 95\% CL excluded region from $\sqrt s=8$ TeV LHC (red, shaded) and  90\% CL excluded region from current $0\nu\beta\beta$ searches (purple, solid), as well as the FCNC constraint on $W_R$ mass (magenta, dotted, vertical).  The future limits from $\sqrt s=14$ TeV LHC (red, dotted), LHeC without (brown, solid) and with (brown, dot-dashed) polarization, as well as from a ton-scale $0\nu\beta\beta$ experiment (purple,dashed) are also shown. }
\label{fig:3}
\end{figure}

For the large mixing case, $e^+e^-\to NN$ may not be the best discovery mode for heavy neutrinos, since the single heavy neutrino production mode $e^+e^-\to N\nu$ is kinematically favored and has a larger cross section~\cite{Banerjee:2015gca}. However, for the small mixing case, naturally arising in the minimal TeV-scale L-R seesaw models, we find that our proposed method gives the best signal sensitivity in most of the kinematically allowed heavy neutrino mass range, as compared to other existing projections from both high and low-energy probes. This is depicted in Fig.~\ref{fig:3}, where we have plotted the 95\% confidence level (CL) sensitivity contours in the $(M_{W_R},M_N)$ plane for our proposed signal, after including all relevant SM backgrounds, as discussed above, and computing the signal-to-background ratio $S/\sqrt{S+B}$, where $S$ ($B$) is the total signal (background) cross section  multiplied by the integrated  luminosity. 

We have considered both $\sqrt s=500$ GeV and 1 TeV options, each with $500~{\rm fb}^{-1}$ integrated luminosity. We find that even without including the enhancement in the cross section due to beam polarization as discussed above, one can still probe $M_{W_R}$ up to 5.3 (7.5) TeV at $\sqrt s=500$ GeV (1 TeV), as shown by the blue (black) solid curves in Fig.~\ref{fig:3}. Including the polarization effects, the sensitivity can be extended to 6.6 (9.4) TeV  at $\sqrt s=500$ GeV (1 TeV), as shown by the blue (black) dashed curves. Note that for small $M_N\lesssim 70$ GeV and depending on the $W_R$ mass, the 3-body decay rate given by Eq.~\eqref{eq:3body} can become so suppressed that there is a non-negligible probability of it giving rise to a displaced vertex~\cite{Castillo-Felisola:2015bha}. To account for this, we have included the probability factor $P=1-\exp{(-L_D/L)}$ in our calculation of the signal, where $L=\gamma \tau_N$ is the decay length of the heavy neutrino with proper lifetime $\tau_N=1/\Gamma_N$ and $\gamma=\sqrt{s}/2M_N$ is the Lorentz boost factor. We have taken $L_D$ to be 1 mm, beyond which we treat it as a displaced vertex signal, which should be analyzed differently~\cite{Castillo-Felisola:2015bha, Antusch:2016vyf}, and dedicated searches at the intensity frontier, such as SHiP~\cite{Alekhin:2015byh} could be more effective than ILC in probing these scenarios.  Due to this reason, and also due to reduced cut efficiencies, our signal sensitivity drops rapidly for very low $M_N$ values. 

To put our results in perspective, we also show in Fig.~\ref{fig:3} the current 95\% CL exclusion (shaded) region from direct searches at the $\sqrt s=8$ TeV LHC (red, solid)~\cite{Aad:2015xaa, Khachatryan:2014dka}. The HL-LHC with $\sqrt s=14$ TeV and 3 ab$^{-1}$ integrated luminosity can probe the region within the red, dotted curve~\cite{Ferrari:2000sp}. We also compare our results with the projected sensitivity~\cite{Mondal:2015zba, Lindner:2016lxq} of an electron-proton collider, such as LHeC or FCC-eh, for an optimistic configuration of 7 TeV proton beam and 140 GeV electron beam, but with the same $500~{\rm fb}^{-1}$ integrated luminosity and 1\% particle misidentification rate as the ILC case, to make a fair comparison. The 95\% CL results with unpolarized electron beam are shown by the brown, solid curve in Fig.~\ref{fig:3}, which can be extended to the brown, dot-dashed curve with +80\% electron beam polarization~\cite{Lindner:2016lxq}. It is clear that the ILC sensitivities derived here transcend both HL-LHC and LHeC projections, as long as the heavy neutrinos are kinematically accessible, simply due to the fact that lepton colliders provide much cleaner environments for new physics searches in the electroweak sector than hadron colliders.  

There are also low-energy constraints on the $W_R$ mass from flavor changing neutral current (FCNC) processes, such as $K-\overline{K}$ and $B_{d,s}-\overline{B}_{d,s}$ oscillations, which effectively rule out $M_{W_R}\lesssim 2.9$ TeV~\cite{Beall:1981ze, Zhang:2007da, Bertolini:2014sua}.  In addition, for the heavy Majorana neutrino coupling to the electron flavor, there are stringent constraints from non-observation of the rare LNV process of neutrinoless double beta decay ($0\nu\beta\beta$)~\cite{Rodejohann:2011mu}. Assuming that the purely RH current contribution to $0\nu\beta\beta$ is the dominant one and using the latest limit from GERDA phase-II on the half-life of $^{76}$Ge, $T_{1/2}^{0\nu}\geq  5.3\times 10^{25}$ yr at 90\% C.L.~\cite{gerda}, one can also exclude a significant portion of the $(M_{W_R},M_N)$ plane~\cite{Tello:2010am, Dev:2013vxa}, as shown by the purple, solid curve in Fig.~\ref{fig:3}. The future ton-scale $0\nu\beta\beta$ experiments with a projected $T_{1/2}^{0\nu}\geq  10^{27}$ yr~\cite{Abgrall:2013rze, Albert:2014afa, Wang:2015raa} can rule out the region to the left of the purple, dashed curve. In any case, the ILC sensitivities are still better than the future limits from $0\nu\beta\beta$ for most of the parameter space shown here.

\section{Conclusion}\label{sec:4}
In conclusion, we reiterate that future $e^+e^-$ colliders provide an unprecedented opportunity for probing TeV-scale seesaw models. In particular, the beam polarization option can be used effectively to distinguish between different contributions to the seesaw signal. We have shown that just measuring the total unpolarized and polarized cross sections of the process $e^+e^-\to NN$ can reveal the nature of the heavy neutrino interaction with the SM sector and probe the heavy-light neutrino mixing parameter. Moreover, applying this method to the minimal TeV-scale L-R seesaw, we can probe an uncharted swath of the parameter space, with the RH gauge boson mass reach up to 6.6 TeV for $\sqrt s=500$ GeV and up to 9.4 TeV for $\sqrt s=1$ TeV. This surpasses the future sensitivities to L-R seesaw as projected by other contemporary proposals, such as HL-LHC and LHeC at the high-energy frontier, and ton-scale $0\nu\beta\beta$ experiments at the low-energy frontier.  Thus, we hope this study will add yet another physics impetus for a future $e^+e^-$ machine. 


\end{document}